\documentclass[preprint, 5p, twocolumn]{elsarticle}

\usepackage{pifont}
\usepackage{natbib}
\usepackage{geometry}
\usepackage{fleqn}
\usepackage{graphicx}
\usepackage{amssymb}
\usepackage{amsmath}
\usepackage{amsfonts}
\usepackage{lineno}
\usepackage{url}

\begin{document}

\title{A new method to calibrate ionospheric pulse dispersion for UHE
  cosmic ray and neutrino detection using the Lunar Cherenkov
  technique}

\author[astron]{R.McFadden\corref{cor1}}
\ead{mcfadden@astron.nl}
\author[csiro]{R.Ekers}
\author[csiro]{P.Roberts}
\cortext[cor1]{Principal corresponding author}
\address[astron]{ASTRON, 7990 AA Dwingeloo, The Netherlands}
\address[csiro]{CSIRO, Astronomy and Space Science, Epping, NSW 1710, Australia}

\begin{abstract}
UHE particle detection using the lunar Cherenkov technique aims to
detect nanosecond pulses of Cherenkov emission which are produced
during UHE cosmic ray and neutrino interactions in the Moon's
regolith. These pulses will reach Earth-based telescopes
dispersed, and therefore reduced in amplitude, due to their
propagation through the Earth's ionosphere. To maximise the
received signal to noise ratio and subsequent chances of pulse
detection, ionospheric dispersion must therefore be corrected, and
since the high time resolution would require excessive data
storage this correction must be made in real time. This requires
an accurate knowledge of the dispersion characteristic which is
parameterised by the instantaneous Total Electron Content (TEC) of
the ionosphere. A new method to calibrate the dispersive effect of
the ionosphere on lunar Cherenkov pulses has been developed for
the LUNASKA lunar Cherenkov experiments. This method exploits
radial symmetries in the distribution of the Moon's polarised
emission to make Faraday rotation measurements in the visibility
domain of synthesis array data (i. e. instantaneously). Faraday
rotation measurements are then combined with geomagnetic field
models to estimate the ionospheric TEC. This method of ionospheric
calibration is particularly attractive for the lunar Cherenkov
technique as it may be used in real time to estimate the
ionospheric TEC along a line-of-sight to the Moon and using the
same radio telescope.
\end{abstract}
\begin{keyword}
UHE neutrino detection \sep Lunar Cherenkov technique \sep
Detectors - telescopes \sep Ionosphere \sep Lunar polarisation
\end{keyword}

\maketitle

\section{Introduction}

The lunar Cherenkov technique \cite{Dagkesamanskii1989} provides a
promising method of UHE neutrino detection since it utilises the
lunar regolith as a detector; which has a far greater volume than
current ground-based detectors. This technique makes use of
Earth-based radio telescopes to detect the coherent Cherenkov
radiation emitted when a UHE neutrino interacts in the outer
layers of the Moon. It was first applied by Hankins, Ekers and
O'Sullivan using the 64-m Parkes radio telescope
\cite{Hankins1996Asearch} and significant limits have been already
been placed on the UHE neutrino flux by several collaborations
\citep{Hankins1996Asearch, Gorham2004Experimental, Beresnyak2005,
James2010MNRAS, NuMoon2010}.

Electromagnetic pulses originating in the lunar surface will be
dispersed when they arrive at Earth-based receivers due to
propagation through the ionosphere which introduces a
frequency-dependent time delay. This dispersion reduces the peak
amplitude of a pulse, however, dedispersion techniques can be used
to recover the full pulse amplitude and consequently increase the
chances of detection. Accurate dedispersion requires an
understanding of the ionospheric dispersion characteristic and
it's effect on radio-wave propagation.

\section{Effects of Ionospheric Dispersion}

The ionosphere is a weakly ionized plasma which is formed by
ultraviolet ionizing radiation from the sun. Due to its
relationship with the sun, the ionosphere's electron density
experiences a strong diurnal cycle and is also dependent on the
season of the year, the current phase of the 11-year solar cycle
and the geometric latitude of observation. The differential
additive delay, caused by pulse dispersion, is parameterised by
the ionospheric TEC (see Equation \ref{eq:Derived3})

\begin{equation} \Delta t= 0.00134 \times
STEC \times (\nu_{\text{lo}}^{-2}-\nu_{\text{hi}}^{-2}),
\label{eq:Derived3}
\end{equation}
where $\Delta t$ is the duration of the dispersed pulse in
seconds, $STEC$ is the Slant Total Electron Content in electrons
per cm$^{2}$ and $\nu_{\text{lo}}$ and $\nu_{\text{hi}}$ are the
receiver bandwidth band edges in Hz.

Cherenkov emission produces a sub-nanosecond pulse and therefore
detection requires gigahertz bandwidths to achieve the high time
resolution  needed to optimse the signal to noise. Due to
excessive data storage requirements, the only way to exploit such
high data rates is to implement real-time dedispersion and
detection algorithms and to store potential events at full
bandwidth for later processing. This requires an accurate
knowledge of the real-time ionospheric TEC.

Ionospheric dispersion also offers some potential experimental
advantages, particularly for single dish experiments which can not
use array timing information to discriminate against RFI. A lunar
pulse will travel a one-way path through the ionosphere and be
dispersed according to the current ionospheric TEC. Conversely,
terrestrial RFI will not be dispersed at all and any Moon-bounce
RFI will travel a two-way path through the ionosphere and be
dispersed according to twice the current ionospheric TEC.
Therefore performing real-time ionospheric dedispersion will
optimise detection for lunar pulses and provide discrimination
against RFI. Dispersion may also be seen to offer an increase in
dynamic range. If triggering is performed on a dedispersed data
stream while the raw data is buffered, any pulse clipping that
occurs in the triggering hardware can be recovered during offline
processing by reconstructing the pulse from the raw, undispersed
data.

\section{Dedispersion Hardware}

Pulse dispersion can be corrected using matched filtering
techniques implemented in either analog or digital technology.
Early LUNASKA experiments made use of the Australia Telescope
Compact Array (ATCA) which consists of six 22-m dish antennas.
Three antennas were fitted with custom-designed hardware for the
neutrino detection experiments and pulse dedispersion was achieved
through the use of innovative new analog dedispersion filters that
employ a method of planar microwave filter design based on inverse
scattering \cite{Roberts1995}.

As the microwave dedispersion filters have a fixed dedispersion
characteristic, an estimate had to be obtained for the TEC which
would minimise errors introduced by temporal ionospheric
fluctuations. The ATCA detection experiments were performed in
2007 and 2008 during solar minimum and therefore relatively stable
ionospheric conditions. Initial observations were during the
nights of May 5, 6 and 7, 2007 and these dates were chosen to
ensure that the Moon was at high elevation (particularly during
the night-time hours of ionospheric stability) and positioned such
that the ATCA would be sensitive to UHE particles from the
galactic center. The filter design was based on predictions made
using dual-frequency GPS data and assumed a differential delay of
5 ns across the 1.2--1.8 GHz bandwidth. Data available post
experiment revealed that the average differential delay for these
nights was actually 4.39 ns, with a standard deviation of 1.52 ns.

The ionosphere experiences both temporal and spatial fluctuations
in TEC and therefore some signal loss is expected with a fixed
dedispersion filter. A promising digital solution to overcome
these losses lies in the use of high speed Field Programmable Gate
Arrays (FPGAs). An FPGA implementation allows the dedispersion
characteristic to be tuned in real time to reflect temporal
changes in the ionospheric TEC. A fully coherent or predetection
dedispsersion method was pioneered by Hankins and Rickett
\cite{Hankins1971, Hankins1975} which completely eliminates the
effect of dispersive smearing. This is achieved by passing the
predetected signal through an inverse ionosphere filter which can
be implemented in either the time domain, as an FIR filter, or in
the frequency domain.

In 2009, the LUNASKA collaboration started a series of UHE
neutrino detection experiments using the 64-m Parkes radio
telescope. For these experiments, dedispersion was achieved via a
suite of FIR filters implemented on a Vertex 4 FPGA. As GPS TEC
estimates are currently not available in real-time, near real-time
TEC measurements were derived from foF2 ionosonde measurements.
Ionosondes probe the peak transmission frequency (fo) through the
F2-layer of the ionosphere which is related to the ionospheric TEC
squared. A comparison to GPS data available post-experiment
revealed that the foF2-derived TEC data consistently
underestimated the GPS TEC measurements. This is attributed to the
ground-based ionosondes probing mainly the lower ionospheric
layers and not properly measuring TEC contribution from the
plasmasphere \cite{Titheridge1972determination}.

\section{Monitoring the Ionospheric TEC}

Coherent pulse dedispersion requires an accurate knowledge of the
ionospheric dispersion characteristic which is parameterised by
the instantaneous ionospheric TEC.

TEC Measurements can be derived from dual-frequency GPS signals
and are available online from NASA's CDDIS \cite{CDDIS}, however,
these values are not available in real time. The CDDIS TEC data is
sampled at two-hour intervals and is currently published after at
least a few days delay. Estimates derived from foF2 ionosonde
measurements are available hourly from the Australian Ionospheric
Prediction Service \cite{IPS}. However, as discussed, there are
known inaccuracies in the derivation of the foF2-based TEC
estimates.

\begin{figure}[!tbph]
\begin{center}
\includegraphics [width=0.4\textwidth,keepaspectratio]{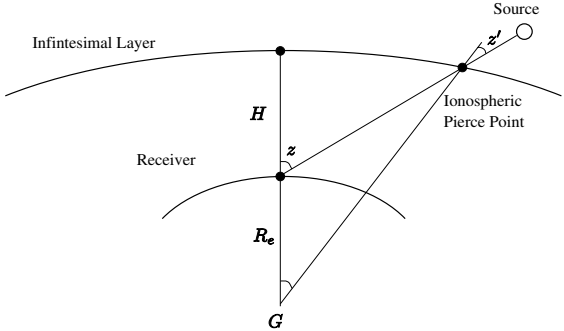}
\caption{Parameters of the Ionospheric Single Layer
Model.}\label{Iono_model}
\end{center}
\end{figure}

Both of these products are published as vertical TEC (VTEC) maps
which must be converted to Slant TEC (STEC) estimates to obtain
the true total electron content through the slant angle
line-of-sight to the Moon. To perform this conversion, the
ionosphere can be modeled as a Single Layer Model (SLM)
\cite{Todorova2008} which assumes all free electrons are
concentrated in an infinitesimally thin shell and removes the need
for integration through the ionosphere. Slant and vertical TEC are
related via

\begin{equation}
STEC=F(z)VTEC. \label{Eq:slant}
\end{equation}
where $F(z)$ is a slant angle factor defined as

\begin{align}
F(z)&=\frac{1}{\cos(z^{\prime})}\\
&=\frac{1}{\sqrt{1-\left(\frac{R_e}{R_e+H}\sin(z)\right)^2}},
\end{align}
$R_e$ is the radius of the Earth, $z$ is the zenith angle to the
source and $H$ is the height of the idealised layer above the
Earth's surface (see Figure \ref{Iono_model}). The CDDIS also use
an SLM ionosphere for GPS interpolation algorithms and assume a
mean ionospheric height of 450 km.

\section{A New Method of Ionospheric Calibration}

As the solar cycle enters a more active phase accurate pulse
dedispersion is becoming a more important experimental concern for
the lunar Cherenkov technique. This requires methods of obtaining
more accurate measurements of the ionospheric TEC.

A new technique has been formulated to obtain TEC measurements
that are both instantaneous and line-of-sight to the Moon. The
ionospheric TEC can be deduced from Faraday rotation measurements
of a polarised source combined with geomagnetic field models,
which are more stable than ionospheric models (the CCDIS
\cite{CDDIS} states that ionospheric TEC values are accurate to
$\sim$20\% while the IGRF \cite{IGRF} magnetic field values are
accurate to better than 0.01\%). Lunar thermal emission can be
used as the polarised source since Brewster angle effects produce
a nett polarisation excess in the emission from the lunar limb
\cite{Heiles1963}. This provides a method for calibrating the
ionosphere directly line-of-sight to the Moon and makes the lunar
Cherenkov technique extremely attractive for UHE cosmic ray and
neutrino astronomy as it allows the characteristic dispersion to
be used as a powerful discriminant against terrestrial RFI whilst
removing the need to search in dispersion-space.

The unique constraints of an UHE neutrino detection experiment
using the lunar Cherenkov technique conflict with traditional
methods of planetary synthesis imaging and polarimetry which
requires a complete set of spacings and enough observing time for
earth rotation. Therefore to apply this method of ionospheric
calibration to the ATCA detection experiments, innovations in the
analysis of lunar polarisation observations were required. In
particular, a method of obtaining lunar Faraday rotation estimates
in the visibility domain (i. e. without Fourier inversion to the
image plane) had to be developed.

Working in the visibility domain removes both the imaging
requirement of a compact array configuration, which would increase
the amount of correlated lunar noise between receivers, and also
removes the need for earth rotation allowing measurements to
obtained in real time. This technique makes use of the angular
symmetry in planetary polarisation distribution. The intrinsic
thermal radiation of a planetary object appears increasingly
polarised toward the limb, when viewed from a distant point such
as Earth \cite{Heiles1963, SPORT2002}. The polarised emission is
radially aligned and is due to the changing angle of the planetary
surface toward the limb combined with Brewster angle effects. The
angular symmetry of this distribution can be exploited by an
interferometer so that an angular spatial filtering technique may
be used to obtain real-time position angle measurements directly
in the visibility domain. The measured position angles are
uniquely related to the corresponding $uv$ angle at the time of
the observation. Comparison with the expected radial position
angles, given the current $uv$ angle of the observation, gives an
estimate of the Faraday rotation induced on the Moon's polarised
emission. Faraday rotation estimates can be combined with
geomagnetic field models to determine the associated ionospheric
TEC and subsequently provide a method of calibrating the current
atmospheric effects on potential Cherenkov pulses.

Observations of the Moon were taken using the 22-m telescopes of
the Australia Telescope Compact Array with a center frequency of
1384 MHz. At this frequency the Moon is in the near field of the
array, however, investigation of the Fresnel factor in polar
coordinates showed that it has no dependence on the spatial
parameter, which determines the polarisation distribution of a
planetary body.

Using the angular spatial filtering technique, position angle
estimates were calculated directly in the visibility domain of the
lunar observational data. The Faraday rotation estimates were
obtained by comparing these angles to the instantaneous $uv$ angle
and the resultant Faraday rotation estimates were averaged over
small time increments to smooth out noise-like fluctuations. Since
the polarised lunar emission received on each baseline varied in
intensity over time, there were nulls during which the obtained
position angle information was not meaningful. A threshold was
applied to remove position angle measurements taken during these
periods of low polarised intensity and baseline averaging was
considered necessary as the results on each baseline were slightly
different and each affected differently by intensity nulls. The
Faraday rotation estimates were converted to estimates of
ionospheric TEC via

\begin{equation}
\Omega = 2.36 \times 10^4 \nu^{-2}\int_{\text{path}}N(s)B(s)\cos
(\theta) ds, \label{eq:Faraday_rotation}
\end{equation} where $\Omega$ is the rotation angle in radians, $f$ is the signal
frequency in Hz, $N$ is the electron density in m$^3$, $B$ is the
geomagnetic field strength in T, $\theta$ is the angle between the
direction of propagation and the magnetic field vector and $ds$ is
a path element in m.

\begin{figure}
\begin{center}
\includegraphics [width=0.5\textwidth,keepaspectratio]{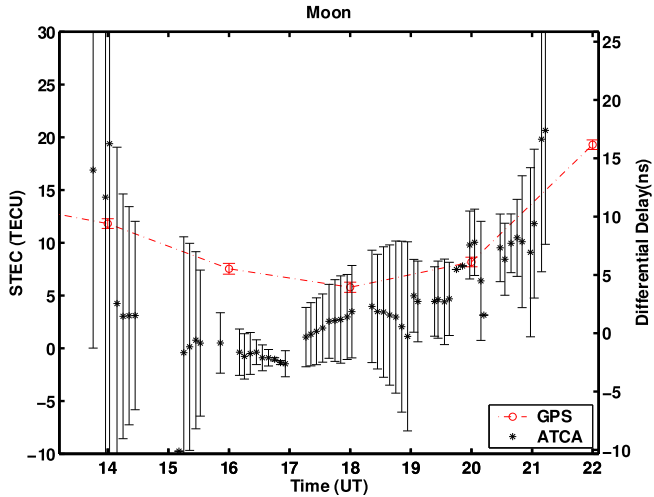}
\caption{Lunar Faraday rotation estimates converted to
(\emph{left}) ionospheric TEC values and (\emph{right}) the
differential delay across 1.2--1.8 GHz.}\label{TECdelay}
\end{center}
\end{figure}

To evaluate the effectiveness of this new ionospheric calibration
technique, the TEC results were compared against ionospheric TEC
estimates derived from dual-frequency GPS data (Figure
\ref{TECdelay}) . Slant angle factors were used to convert the GPS
VTEC estimates to STEC toward the Moon for comparison with the
ATCA data. Both data sets exhibited a similar general trend of
symmetry around the Moon's transit. However, the ATCA data often
underestimated the GPS data, particularly around 14:30--17:00 UT
where the STEC estimates may have been influenced by bad data from
the shorter baselines or due to TEC contribution from the
plasmasphere which is not in the presence of a magnetic field
\cite{Titheridge1972determination}. These observations were taken
when the TEC was very low and therefore the relative error in the
TEC estimates is large.

\section{Conclusions}

As the sun enters a more active phase, accurate ionospheric pulse
dispersion is becoming a more important experimental concern for
UHE neutrino detection using the lunar Cherenkov technique.
Hardware dedispersion options rely on the accuracy of real-time
ionospheric TEC measurements and, while there are a few options
available for obtaining these measurements, they are not currently
available in real time nor directly line-of-sight to the Moon. A
new ionospheric calibration technique has been developed. This
technique uses Faraday rotation measurements of the polarised
thermal radio emission from the lunar limb combined with
geomagnetic field models to obtain estimates of the ionospheric
TEC which are both instantaneous and line-of-sight to the Moon.
STEC estimates obtained using this technique have been compared to
dual-frequency GPS data. Both data sets exhibited similar features
which can be attributed to ionospheric events, however, more
observations are required to investigate this technique further.

\section{Acknowledgements}

This research was supported as a Discovery Project by the
Australian Research Council. The Compact Array and Parkes
Observatory are part of the Australia Telescope which is funded by
the Commonwealth of Australia for operation as a National Facility
managed by CSIRO.

\biboptions{square,sort&compress}
\bibliographystyle{elsarticle-num}
\bibliography{My_bib2}

\end{document}